\documentclass[a4paper,12pt]{article}
\usepackage{amsmath,amssymb,graphicx,setspace,epsfig,color,footmisc}
\usepackage{bm}

\def \Re{{\rm I\kern -1.6pt{\rm R}}}
\def \Expect{{\rm I\kern -1.6pt{\rm E}}}
\def\keywords{\vspace{.5em}
{\textit{Keywords}:\,\relax%
}}
\def\endkeywords{\par}

\begin{document}

\title{Parameter Estimation Through Ignorance}

\author{Hailiang Du$^{1}$\quad Leonard A. Smith$^{1,2}$  \\[2ex]
        $^1$Centre for the Analysis of Time Series,\\
       London School of Economics, London, UK\\[1ex]
   $^2$Pembroke College, Oxford, UK}


\date{\today}

\maketitle

\begin{abstract}

Dynamical modelling lies at the heart of our understanding of physical systems. Its role in science is deeper than mere operational forecasting, in that it allows us to evaluate the adequacy of the mathematical structure of our models. Despite the importance of model parameters, there is no general method of parameter estimation outside linear systems. A new relatively simple method of parameter estimation for nonlinear systems is presented, based on variations in the accuracy of probability forecasts. It is illustrated on the Logistic Map, the Henon Map and the 12-D Lorenz96 flow, and its ability to outperform linear least squares in these systems is explored at various noise levels and sampling rates. As expected, it is more effective when the forecast error distributions are non-Gaussian. The new method selects parameter values by minimizing a proper, local skill score for continuous probability forecasts as a function of the parameter values. This new approach is easier to implement in practice than alternative nonlinear methods based on the geometry of attractors or the ability of the model to shadow the observations. New direct measures of inadequacy in the model, the ``Implied Ignorance'' and the information deficit are introduced.


\end{abstract}

\endkeywords

\vspace{1cm}

The estimation of physical constants (parameters) plays a central role in the physical sciences. Yet there is no general method of parameter estimation for nonlinear dynamical systems~\cite{Sornette}. In what may have been one early use of least squares (see~\cite{Stigler,Teets}), Gauss~\cite{Gauss} predicted where the newly discovered Ceres would appear as it emerged from behind the sun. The prediction involved both a parameter (Newton's gravitational constant) and initial conditions (brief observations of Ceres before occultation). This success, where other methods failed,  supported both the least squares approach and the mathematical form of Newton's Laws. A new Minimum Ignorance (MI) approach to parameter estimation for use in dynamical systems is presented and illustrated in nonlinear cases where the common ``least squares'' approaches can be systematically biased.  While the focus is  on cases in which the data archive is relatively large and the model structure is correct~\footnote{Specifically, cases where there is a parameter value for which the mathematical model is empirically adequate. The motion of Mercury is inconsistent with the mathematical form of Newton's Laws, an internally consistent description requires General Relativity. It is not that the value of a parameter in Newton's Laws is uncertain, but rather the value is indeterminate~\cite{Stern}: no value will yield results consistent with observed planetary motion. In the ``Perfect Model Scenario'', the ``True'' parameter is unknown, but does exist.~\cite{Chaosbook}}, the approach may prove useful outside this Perfect Model Scenario~\cite{Smith02}. In the MI approach, many large sets of probability forecasts are made, each using different parameter values; the quality of those parameter values is determined by the quality of the corresponding set of probability forecasts.  A measure of the internal consistency of probability forecasts is also introduced, providing quantitative insight into modelling inadequacy.

Parameter estimation  for deterministic nonlinear models poses several challenges, as nonlinear processes can be sensitive to initial conditions and parameter specifications. Traditional methods, like least squares, are sub-optimal when forecast errors are non-Gaussian, even if the observational uncertainties are normally distributed. One aim of this paper is to stress that fact given the common, and often unguarded, use of least squares. Several methods have been proposed to address the shortcomings of traditional methods: McSharry and Smith~\cite{mcsharrys99} estimate model parameters by incorporating the global behaviour of the model into the selection criteria; Creveling et al~\cite{Crevel}, Maybhate and Amritkar~\cite{Maybhate} have exploited synchronisation for parameter estimation; Smith et al.~\cite{IS3} focused on the geometric properties of trajectories; Heald and Stark~\cite{Heald} include estimation of the noise model. Recently Quinn and Abarbanel~\cite{Abarbanel} have demonstrated parameter and state can be estimated via evaluation of a discrete time path integral in model state space. They also note applications in a number of fields including neurobiology, atmospheric and oceanic sciences, cell biology, chemical engineering, wastewater treatment and biochemistry. There are also variational approaches~\cite{Schittkowski}, multiple shooting methods~\cite{Voss} and sequential methods based loosely on the Kalman Filter~\cite{Voss,Dempster,Ghahramani,Ghil}. Several of these alternative approaches are contrasted with the MI approach in the conclusion section.

Results of MI parameter estimation are presented and critically examined  for three chaotic systems: the Logistic Map, the Henon Map and the 12-D Lorenz96 flow. MI is shown to outperform linear least squares in these systems when the nonlinearities are relevant; at small lead times and low noise levels MI and LS are comparable. The MI method does not solve the problem of nonlinear parameter estimation completely, but it does highlight the failure of common linear methods and allow significant progress in some nonlinear cases, progress which may generalize beyond the Perfect Model Scenario.

\begin{center}
{\bf Technical Problem Statement}
\end{center}

Parameter estimation is a ubiquitous problem in scientific modelling.~\cite{Sornette,mcsharrys99,Crevel,IS3,Heald,Abarbanel,Bos,Bollt} While well understood in linear systems~\cite{LSE,Inference,TLSE}, challenges remain in nonlinear systems~\cite{Taran}. Discussions of parameter estimation typically assume: dynamical systems are linear (or can be linearised), the mathematical structure of the model is perfect (thus ``True'' parameter values exist) and that the statistics of observational uncertainty are known (the ``noise model'' is perfect). A more complete discussion is provided by Tarantola~\cite{Taran} who in Figure 3.2 sketches six schematic examples, four that are linear or linearisable, one requiring fully nonlinear methods, and one too complex for his methods to be used. Problems of the fifth category in the context of prediction, the so-called ``forward problem'', are approached here.

Assume the evolution of a system state $\textbf{x}_{i}\in \mathbb{R}^{m}$ is governed by finite dimensional,
discrete time, deterministic nonlinear dynamical system:
\begin{eqnarray}
 \label{eq:model}
 \textbf{x}_{i+1}=F(\textbf{x}_{i},\textbf{a}),
\end{eqnarray}
where $\textbf{x}\in \mathbb{R}^{m}$ and the model's parameters are contained in the vector $\textbf{a}\in \mathbb{R}^{l}$.
For $m=1$, the state $x_{i}$ is a scalar. For simplicity forecasts are evaluated on a scalar observation below, even when $m>1$. Assuming additive measurement noise $\delta_{i}$ yields
observations $s_{i}=x_{i}+\delta_{i}$. A~ set of $l+1$ sequential measurements
$s_{i},s_{i+1},...,s_{i+l}$ would, in general, be sufficient to determine $\textbf{a}$ in a noise free setting
(i.e. $\delta_{i}=0$ $\forall i$)~\cite{mcsharrys99}. With noise, the task is somewhat harder.

Given model structure $F(x,\textbf{a})$ and observations generated by particular parameter $\textbf{a}_{0}$
(the ``True'' parameter value), one can identify values for $\textbf{a}$ consistent with the available information. Parameter estimates are made on the basis of the skill of the probability forecast. To ease comparison with previous work~\cite{mcsharrys99,IS3}, the approach is illustrated using 3 nonlinear models: The 1-D Logistic Map:
\begin{eqnarray}
 \label{eq:logmap1}
 F(x,a)=1-ax^2
\end{eqnarray}
and the 2-D Henon Map:
\begin{eqnarray}
 \label{eq:Henon}
 x_{i+1}=1-ax_{i}^{2}+y_{i} \nonumber \\
 y_{i+1}=bx_{i}
\end{eqnarray}
and a 12-D Lorenz96 flow~\cite{Lorenz96}.
\medskip
\begin{center}
{\bf Minimum Ignorance Parameter Estimation}
\end{center}
The least squares (LS) method estimates parameters by
minimising the root mean square error of a point forecast.
Even given infinite data, the optimal LS solution is biased when applied to the Logistic Map~\cite{mcsharrys99}.
The LS method fails because the assumption of Independent Normal Distributed (IND) forecast errors does not hold, even with IND observational noise. This is to be expected in nonlinear models.

A point value based on an imperfectly observed initial state is incomplete as a forecast~\cite{Lorenz65}; given observational uncertainty, an ensemble of initial states
of the system consistent with given observations~\cite{Fermi} is required to propagate this initial uncertainty, suggesting probabilistic forecasts via Monte Carlo ensembles.

\medskip
\centerline{\em Scoring Probabilistic Forecasts}
\medskip
A probabilistic skill score is a function $S(p(y),Y)$, where $Y$ is the outcome and $p(y)$ is a probability forecast~\cite{Adress}. The Ignorance Score~\cite{Good,IgnRS} is given by:
\begin{eqnarray}
 \label{eq:score}
    S(p(y),Y)=-log_{2}(p(Y))
\end{eqnarray}
Ignorance is the only proper local score for continuous variables~\cite{proper, Score}. In practice, given N forecast-outcome pairs $(p_{i}(y),Y_{i}, i=1,...,N)$, the Empirical Ignorance is:
\begin{eqnarray}
 \label{eq:Ign}
    S_{EI}(p(y),Y)=\frac{1}{N}\sum_{i=1}^{N}-log_{2}(p_{i}(Y_{i}))-S_{clim},
\end{eqnarray}
where $S_{clim}$ is defined using the unconditional probability or ``climatology'' of $y$, denoted $p_{c}(y)$;
this is simply the natural measure projected onto the forecast variable. The zero skill of Ignorance is then
\begin{eqnarray}
 \label{eq:Sclim}
    S_{clim}=\int-p_{c}(y)log_{2}(p_{c}(y)) dy
\end{eqnarray}
While other proper skill scores might be used in this context, Ignorance is the only proper local skill score for continuous variables; it is invariant under smooth changes of coordinates. Insuring these properties is desirable in parameter estimation.

\medskip
\centerline{\em Ensembles and Probability Forecasting}
\medskip
An ensemble forecast is based on a collection of simulations simultaneously. There are many methods for forming an ensemble of initial states~\cite{EnKF94,IS1,Palmer}.
Perhaps the simplest method is to add draws from the inverse of the observational noise to the observation to define ensemble members. In that case ensemble members are equally weighted, as each ensemble member is
an independent draw.
With this Inverse Noise method, the initial states are unlikely to be consistent with the long term model dynamics (e.g. they are not ``on the attractor'' should one exist).

Continuous forecast distributions can be produced from an ensemble by kernel dressing its members. Standard kernel dressing is used below (see~\cite{Adress,IgnRS} for more details, and~\cite{Raftery} for a Bayesian approach). Define an $N_{e}$ member ensemble at time $i$ to be $X_{i}=[x^{1}_{i},...,x^{N_{e}}_{i}]$ and treat all ensemble members as exchangeable: the ensemble interpretation methods used do not depend on the ordering of the ensemble members~\cite{Adress}. Standard kernel dressing transforms the ensemble members into a probability density function $p_{m}$ where:
\begin{eqnarray}
 \label{eq:SKD}
 p_{m}(y:X,\kappa)=\frac{1}{N_{e}\kappa}\sum^{N_{e}}_{j=1}K\left(\frac{y-x^{j}}{\kappa}\right).
\end{eqnarray}
In this case the forecast distribution is a sum of Gaussian kernels $K(\cdot)$, the $j^{th}$ ensemble member being replaced by kernel centred at $x^{j}$. For each value of $\textbf{a}$, the kernel width, $\kappa$, is chosen to minimise the Empirical Ignorance defined in equation~\ref{eq:Ign} above~\cite{Adress}. There remains the chance that the verification lies outside the range of any finite ensemble, even if the verification is selected from the same distribution as the ensemble itself; the probability of this happening is
$\gtrsim\frac{2}{N_{e}}$. Given the nonlinearity of the model, these points may be very far from the ensemble, and appear as ``outliers'' or ``bad busts''.

Given a sample climatology of the system from historical data,
probabilistic forecasts may be improved out-of-sample by blending the dressed ensemble
with the sample climatology~\cite{Adress}, thereby allowing both narrower kernels and fewer bad busts.
Blending with climatology yields the forecast distribution:
\begin{eqnarray}
 \label{eq:blending}
    p(y)=\alpha p_{m}(y)+(1-\alpha)p_{c}(y)
\end{eqnarray}
where $p_{m}$ reflects the ensemble and $p_{c}$ the climatology. The probability forecast obtained will be a function of $\textbf{a}$. Values of $\textbf{a}$ with small Empirical Ignorance are deemed better.

Comparing forecast performance of different models is not a \emph{fair} comparison without blending climatology. It might be the case that, without blending climatology Model A outscores Model B, while after blending climatology Model B scores higher than Model A. Since the sample climatology is available to any model, the comparison should include this information.

\begin{center}
{\bf Evaluation and Results}
\end{center}

\begin{figure}[h!]
\centering
  \vbox{
  \epsfig{file=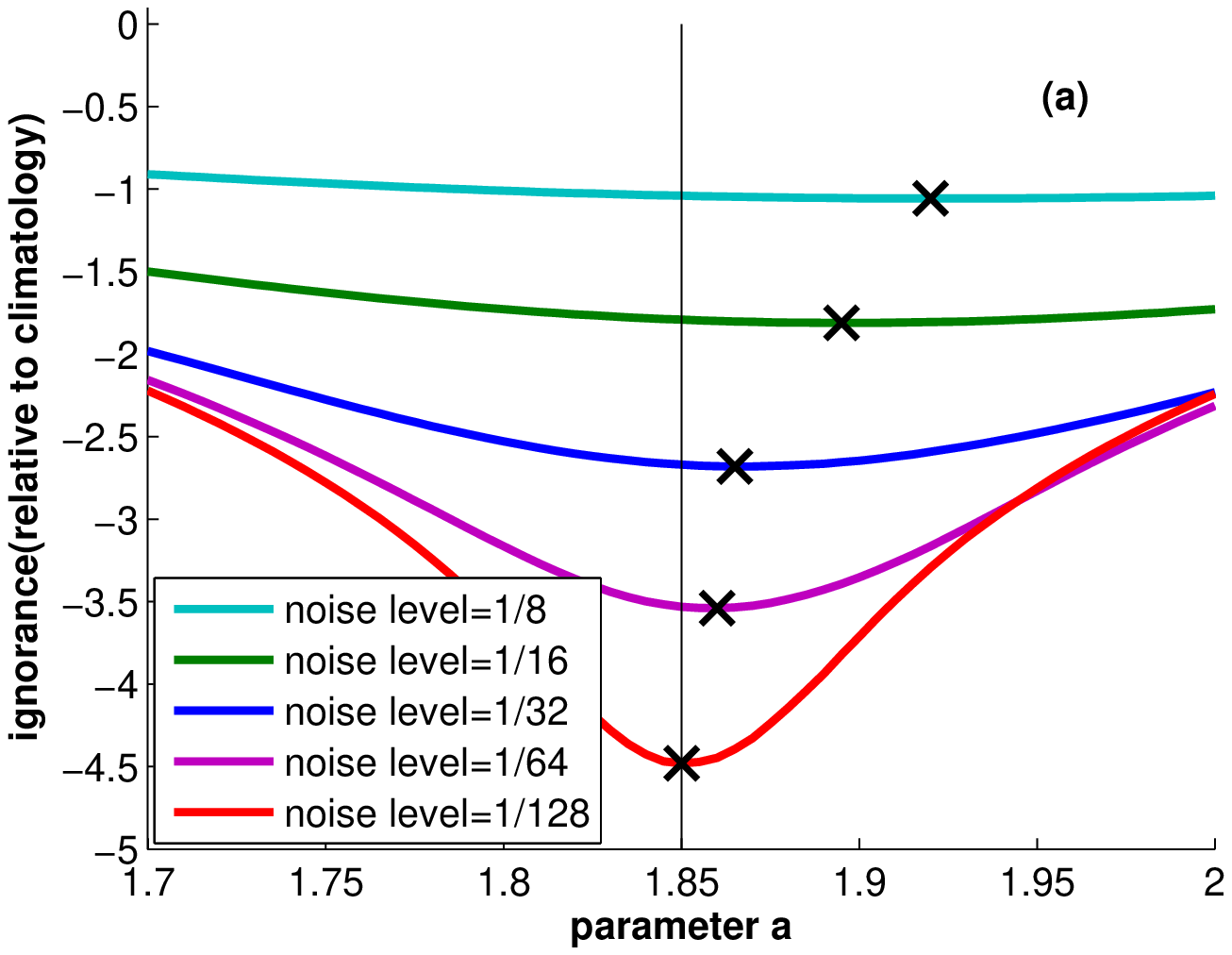, width=6.5cm}
  \epsfig{file=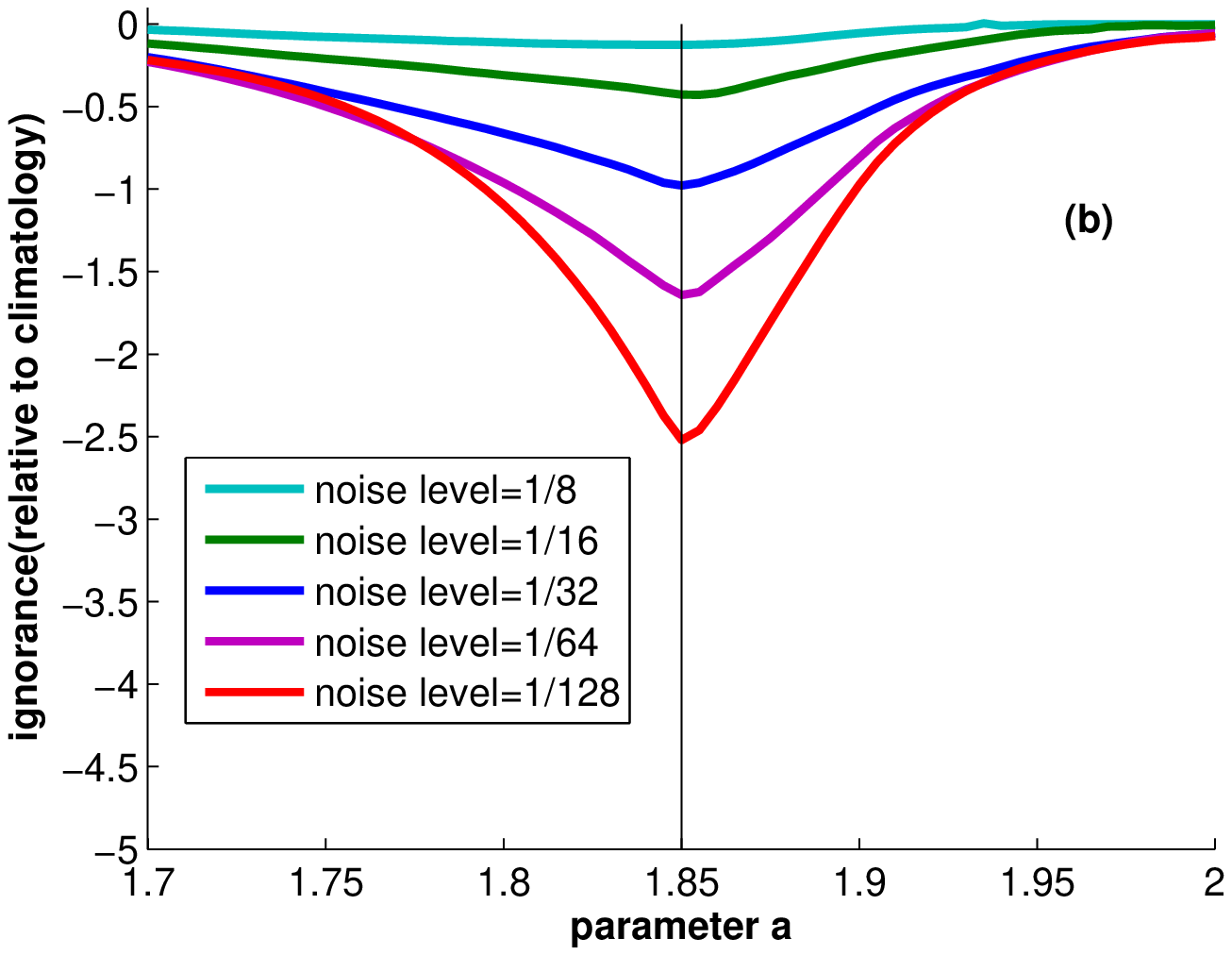, width=6.5cm}
}
\caption{(Color) Minimum Ignorance parameter estimation for Logistic Map with \mbox{$\textbf{a}=1.85$}; initial condition ensembles are formed by Inverse Noise. Five different noise levels are tested, each given 1024 forecasts; (a) Ignorance as a function of $\textbf{a}$ for $\tau=1$, the minima are marked with an ``x''; (b) for $\tau=4$.}

  \label{fig:logisticest}
\end{figure}

\begin{figure}[h!]
\centering
  \vbox{
  \epsfig{file=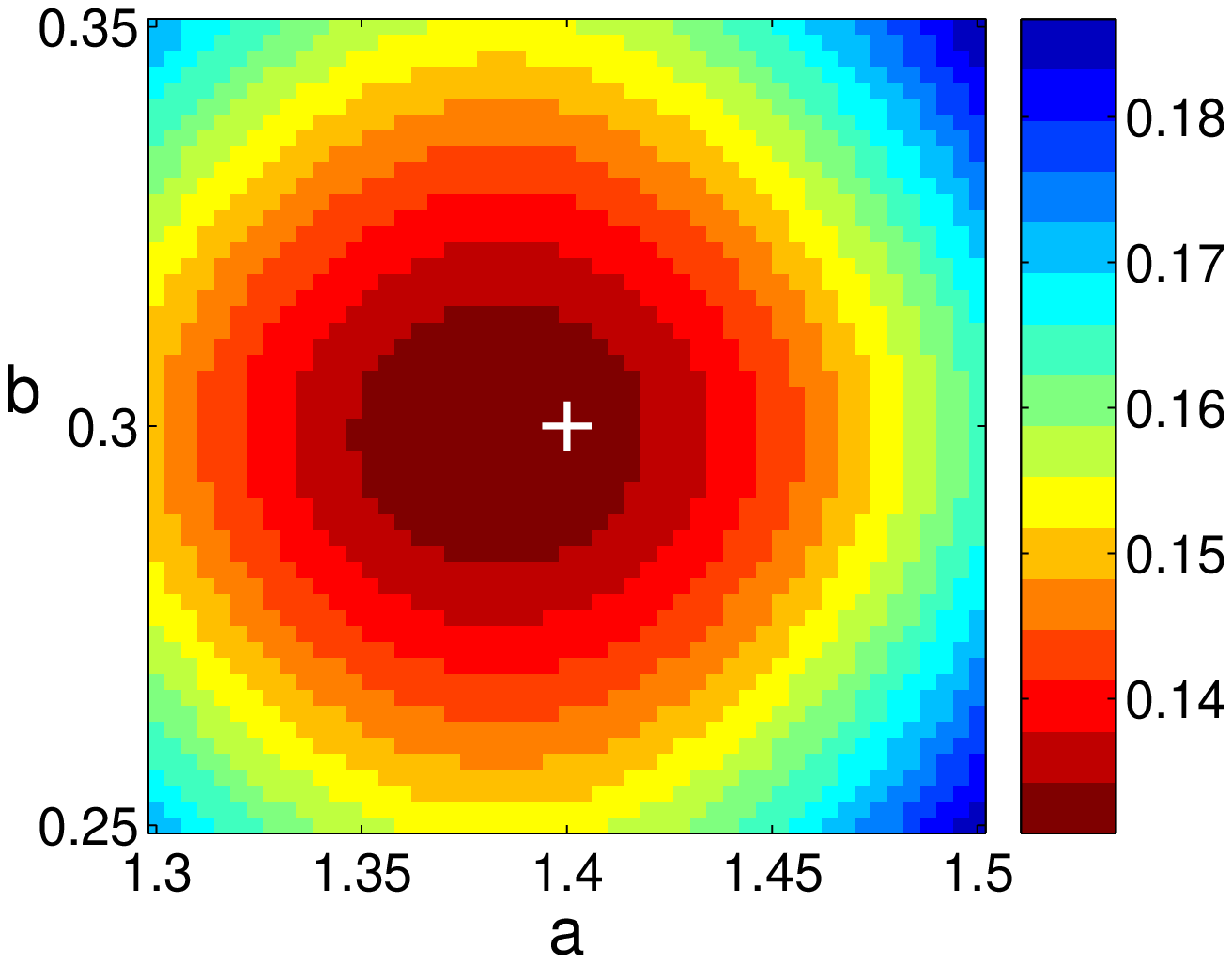, width=6.5cm}
  \epsfig{file=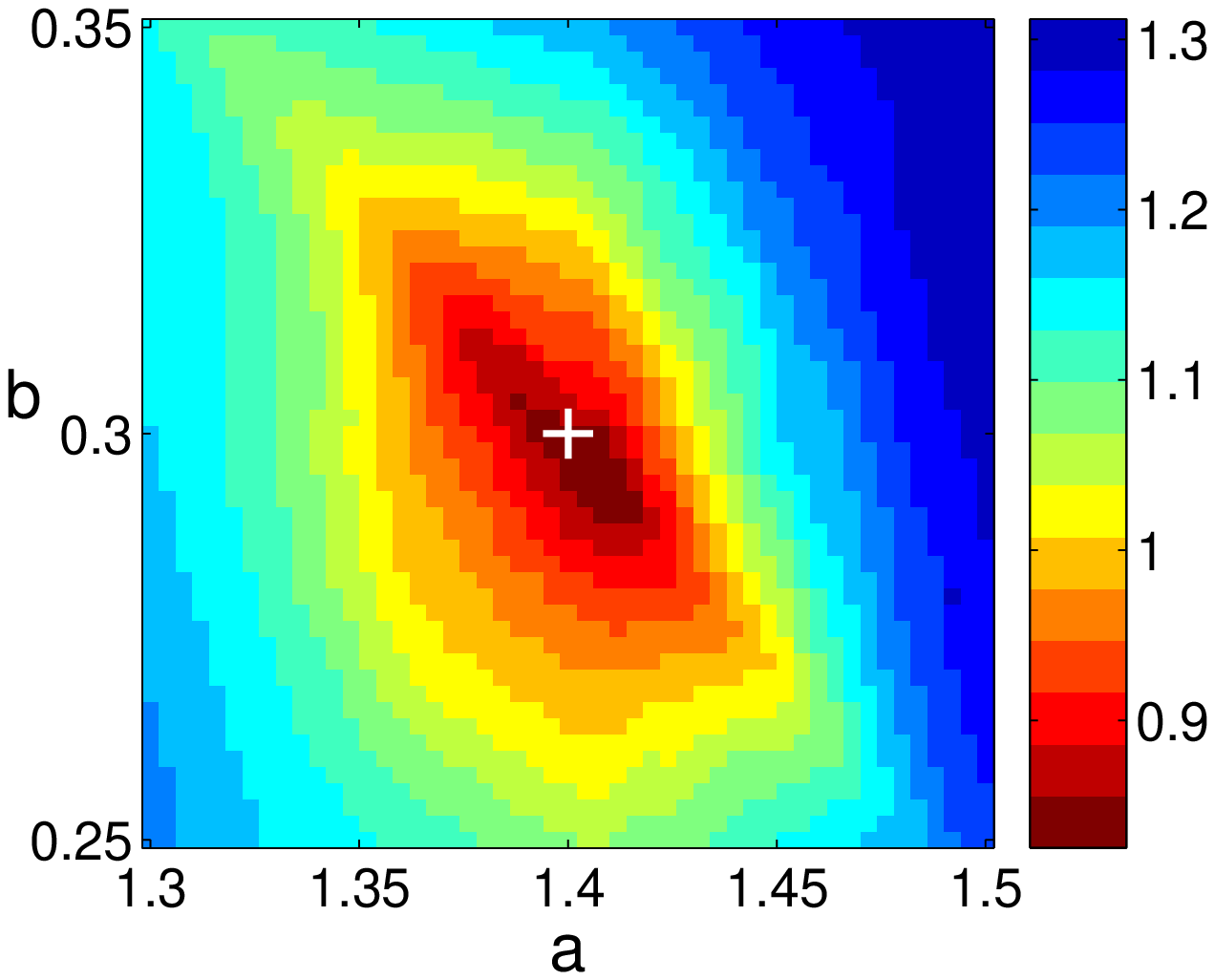, width=6.5cm}
}
  \caption{(Color) Parameter estimation for Henon map (a=1.4; b=0.3), noise level=0.05, given 1024 forecasts at lead time 4. (a) a cost function based on LS, (b) a cost function based on forecast Ignorance. }
  \label{fig:henonest}
\end{figure}

Figures~\ref{fig:logisticest} and~\ref{fig:henonest} show the Empirical Ignorance scores as a function of lead time, $\tau$, and parameter value $a$ for the Logistic Map and for the Henon Map.
Figure~\ref{fig:logisticest} shows five different noise levels $\sigma$, for two lead times. In panel (a) $\tau=1$ and panel (b) $\tau=4$.
The vertical line marks the ``True'' parameter value of 1.85, Figure~\ref{fig:henonest} reports results from the Henon map showing how MI approach outperforms a LS method. The '+' in each panel reflects the ``True'' values of $a$ and $b$. Panel (a) shows the inferiority of the LS error.

\begin{figure}[h]
  \centering
\vbox{
  \epsfig{file=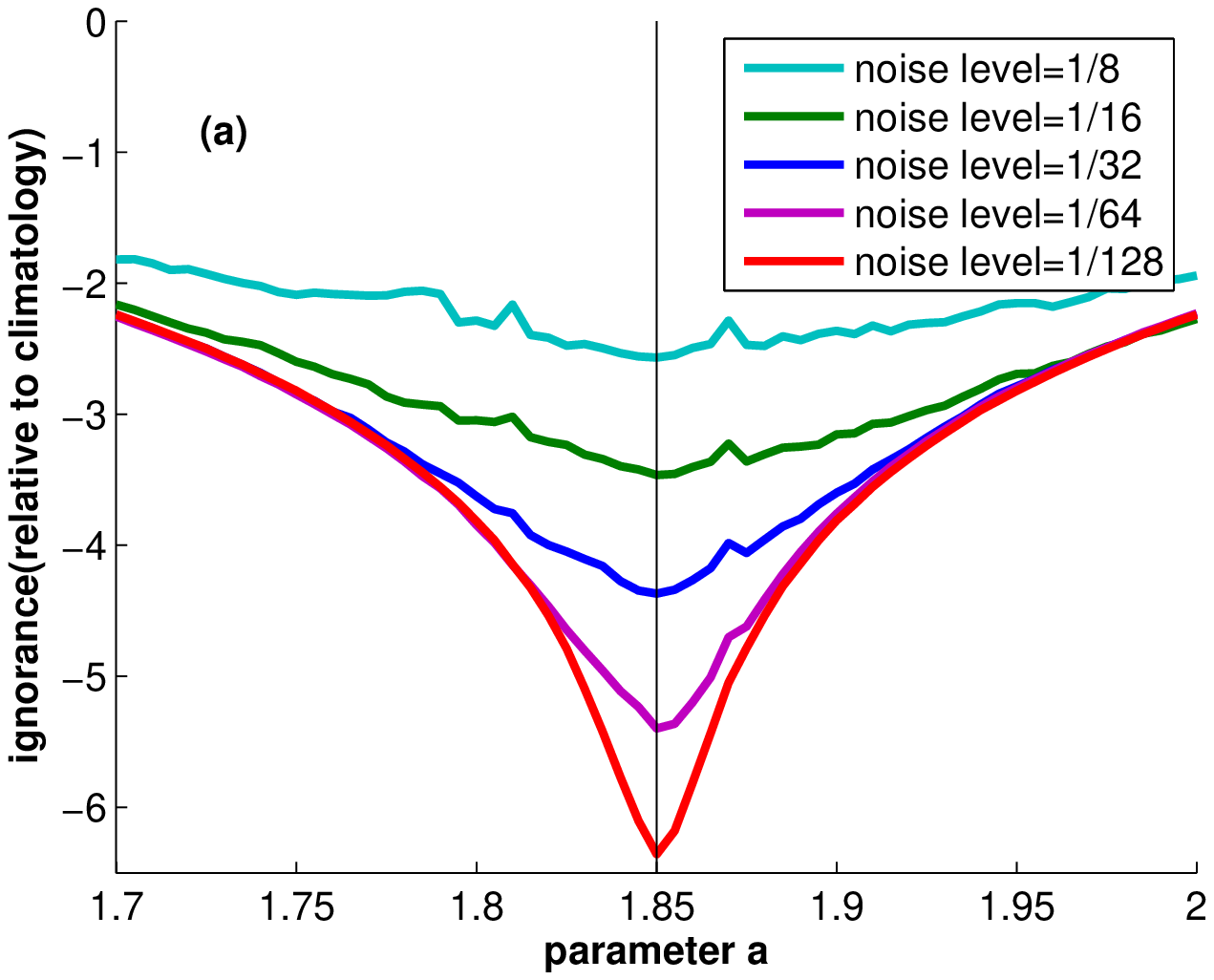, width=6.5cm}
  \epsfig{file=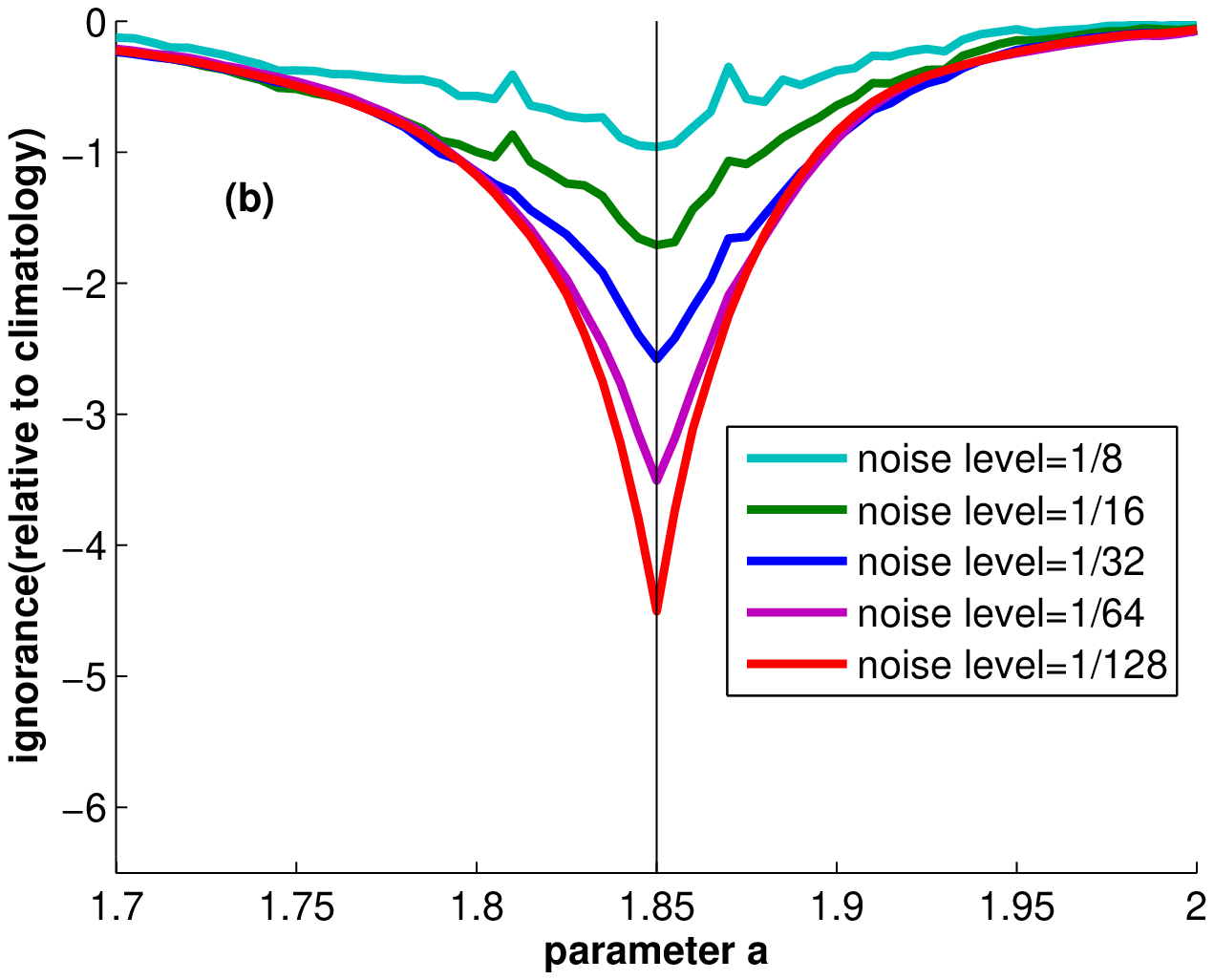, width=6.5cm}
}
  \caption{(Color online) Parameter estimation for Logistic Map with $\textbf{a}=1.85$ using dynamically consistent ensembles. Contrast the (improved) Ignorance value relative to Figure~\ref{fig:logisticest} where the same lead times and noise level are used.}
  \label{fig:perfect ensemble}
\end{figure}

Returning to Figure~\ref{fig:logisticest}(a), note the bias away from the ``True'' value. MI estimates at longer $\tau$ (Figure~\ref{fig:logisticest}(b)) tend to provide less biased estimates. The small $\tau$ bias is due to imperfections in the initial ensemble: neither the observation itself nor the initial ensemble formed by Inverse Noise are consistent with the long time dynamics. The natural measure of the Logistic Map is not uniform; for some parameter values it may be fractal. A dynamically consistent ensemble is an ensemble of initial conditions which are not only consistent with the observational noise, but also consistent with the natural measure.~\footnote{\label{fn:foot}Note for a structurally perfect model, the dynamically consistent ensemble will approach a perfect ensemble\cite{Fermi} at the ``True'' parameter value when a long window of observations is considered.} Figure~\ref{fig:perfect ensemble} shows using a more dynamically consistent ensemble of initial conditions (in this case merely consistent with the current observation~\footnote{Here the dynamically consistent ensemble of initial conditions are only consistent with the current observations; requiring consistency with a series of observations would result in more informative ensembles. To locate states also consistent with more past observations are much more costly.})  produces less biased results at short $\tau$ and also improves larger $\tau$.



To demonstrate that the advantage of MI parameter estimation need not vanish in higher dimensional systems, the single parameter Lorenz96 system~\cite{Lorenz96} is considered with $m=12$ and the parameter $F=17$, using Inverse Noise ensembles. Nonlinear effects are reduced at smaller lead times $\tau$ and lower noise levels $\sigma$. In Lorenz96, the estimation error of LS and MI are roughly the same with $\tau=0.5$ and $\sigma=0.1$ (this is $\sim 0.2\%$ of the range of the data). Increasing the noise level to $\sigma=1$, the estimation error from LS is $\sim 8$ times that of MI. Alternatively keeping $\sigma=0.1$ and increasing to $\tau=1$ yields an error in the LS estimate $\sim 3$ times larger. For any smooth $F(x)$ the linear approximation will hold in the limit of infinitesimally small observational noise; even in this limit MI estimation will outperform linear methods which, like variants of the Kalman filter fail, to respect the natural measure of $F$.


\begin{center}
{\bf Imperfect Model Scenario}
\end{center}
In the statistics literature, parameters within the Perfect Model Scenario, where a ``True'' value is thought to exist but is unknown, are sometimes referred to as ``quantities with a well defined physical meaning.'' (see for instance,~\cite{Bos}). Here the distinction is made between fitting parameters in a ``physical model'' and a ``curve fitting model'', where in the second case parameters are defined only relative to some goal. As will be demonstrated below, if the mathematical structure of the model is imperfect there is no unique value of the parameter is ``optimal'', the ``best'' parameter may vary with application (the lead time of the forecast, for example). Within Perfect Model Scenario the ``True'' or optimal parameter value exists but is unknown, outside Perfect Model Scenario there it is not unknown but undefined, one is dealing not with uncertainty but ambiguity~\cite{Stern}.

All analysis techniques including LS are limited to exploring the information contained in the data; large forecast-outcome archives and lower observational noise levels contain more information and thus allow better parameter estimates when the model structure is perfect. When the model class does not admit an empirically adequate model, the notion of a ``True'' parameter value is lost. The MI approach remains useful for identifying best parameter in an imperfect model if a notation of ``best'' is defined in terms of forecast performance.

\begin{figure}[h]
  \centering
\vbox{
  \epsfig{file=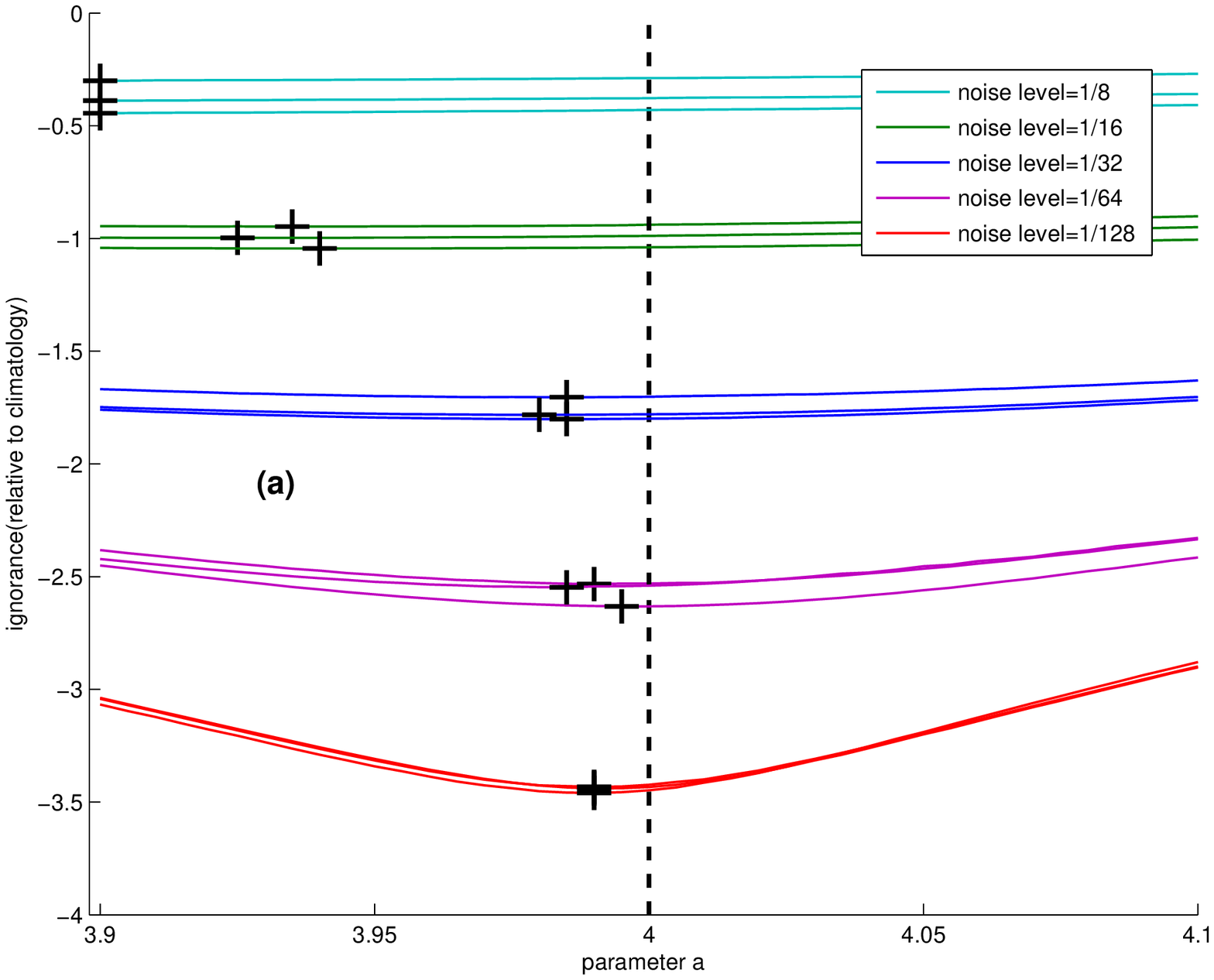, width=6.5cm}
  \epsfig{file=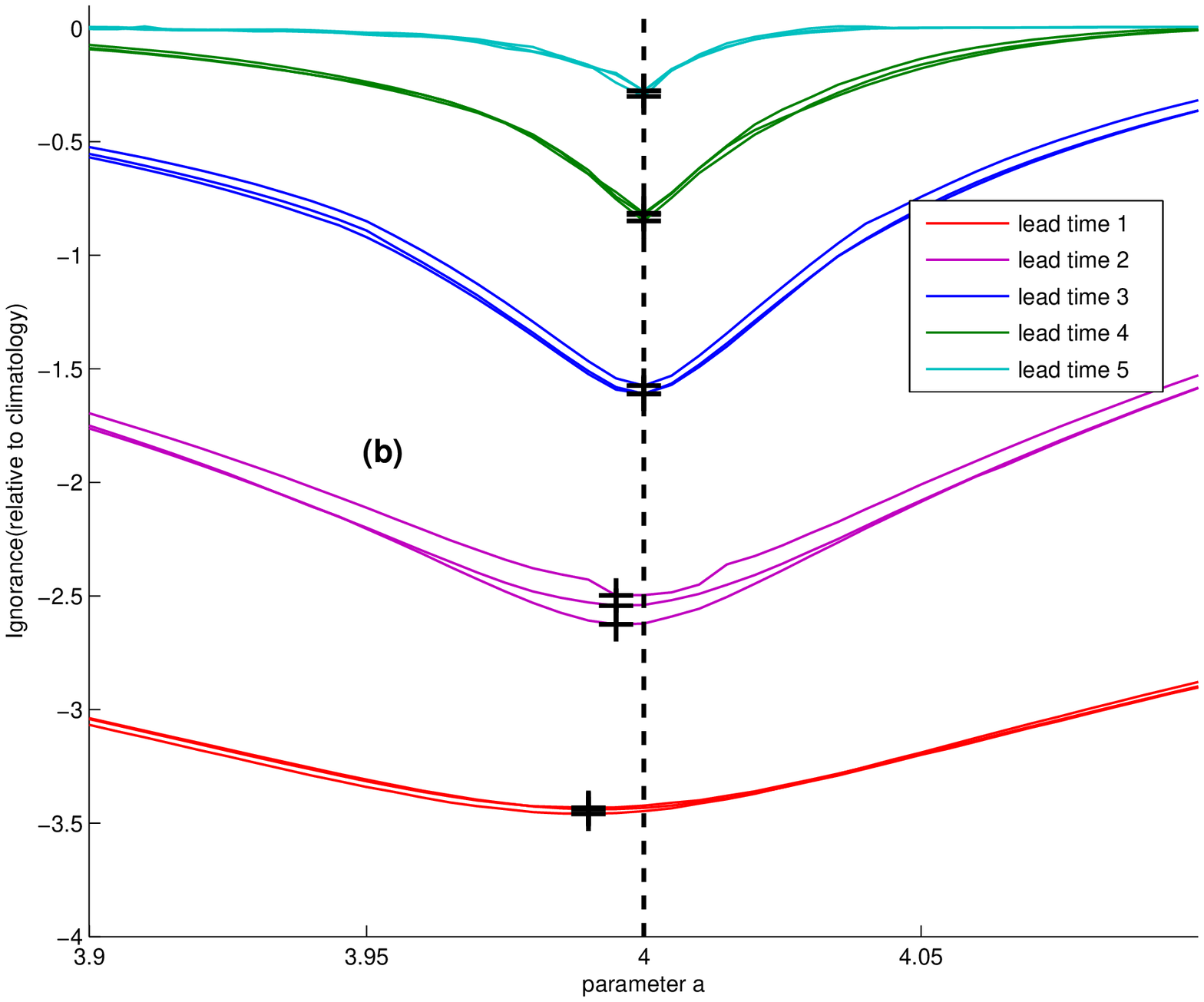, width=6.5cm}
}
  \caption{(Color online) Parameter estimation for Logistic model in the Imperfect Model Scenario, with parameter $\tilde{a}=4$ of Quartic system, using Inverse Noise ensembles. Results from three independent realizations are shown, each given 1024 forecasts; note consistency in locating the minimum ($\times$). The similarity of these three lines indicates the result is robust. a) Empirical Ignorance scores as a function of the parameter value for lead time 1 forecast at several noise level; b) Empirical Ignorance scores as a function of the parameter value and lead time given Noise Level$=1/128$.}
  \label{fig:impms}
\end{figure}

Next consider a system-model pair in the Imperfect Model Scenario. The Quartic system is defined as
\begin{eqnarray}
 \label{eq:qsystem}
	 \tilde{G}(\tilde{x})=\tilde{a}((1-\tilde{\iota} )\tilde{x}(1-\tilde{x})+\frac{4\tilde{\iota}}{5}\tilde{x}(1-2\tilde{x}^2+\tilde{x}^3)).
\end{eqnarray}
The model in this case is
\begin{eqnarray}
 \label{eq:logisticMap}
 G(x)=ax(1-x),
\end{eqnarray}
which is just the Logistic Map in another form~\cite{May}. At $\tilde{\iota}=0$ the model is perfect (as $\tilde{\iota}\rightarrow 0$ the model has structural error); $\tilde{\iota}=0.1$ is considered here. Given the observations generated by the system with additive noise, the goal is to estimate the parameter of the (imperfect) model. Figure~\ref{fig:impms}(a) shows the Empirical Ignorance scores as a function of parameter value for Logistic model at lead time 1, following Figure~\ref{fig:logisticest} five different noise levels are examined. Figure~\ref{fig:impms}(b) the noise level is fixed while 5 different lead times are examined. Note the dashed black line reflects the parameter $\tilde{a}$ used in the Quartic system which need no longer be the target of the model parameter value $a$. Results for three independent experiments are shown, indicating that the bias away from the system parameter value is robust. Figure~\ref{fig:impms} shows the MI estimate varies with lead time and noise level. In both cases the notation of ``best'' is defined in terms of forecast skill given an Inverse Noise initial condition ensemble. The system parameter value $\tilde{a}$ is not equal to the best model parameter value $a$.

\begin{center}
{\bf Further Discussion}
\end{center}

Minimum Ignorance parameter estimation considers the entire forecasting scenario; once the notion of ``best" is defined any alteration of the forecasting scheme may alter the best parameter value. In this section, effects of ensemble formation and kernel dressing are discussed, and an alternative to ``potential predictability" is suggested.

The variance of the standard kernel dressed ensemble is of course always larger
than the variance of the raw ensemble, no matter how the kernel width is actually determined~\cite{Wilks06}.
More complicated dressing methods exist; (Brocker and Smith~\cite{Adress} for example introduced an improved kernel dressing, called ``affine kernel dressing'', that is more flexible and robust).
Standard kernel dressing is used here as it is straightforward to understand, easier to implement, and fit for our purpose. More advanced data assimilation methods may yield more informative ensembles (for example Indistinguishable States~\cite{IS1}, Monte Carlo methods~\cite{Abarbanel,EnKF94}).
If it is costly to run the model (as with weather/climate models), Inverse Noise provides a much faster and cheaper first-pass estimate. There are also alternative low cost distributions one can use to blend with the dressed ensemble forecast
other than the unconditional climatology, for example a dynamical climatology ensemble based on analogues to the current state (see the discussion of eRAP in~\cite{Fermi}). The MI approach generalises beyond estimating ``physical'' parameters as it can be used for structural parameters as, for example, in delay space reconstructions (see Farmer and Sidorowich~\cite{Farmer} and citations thereof) and model reduction~\cite{Bollt07}. Finally, note that it is also possible to estimate the parameters of the noise model(s)~\cite{Heald} within the MI framework.

\begin{figure}[h]
  \centering
\vbox{
  \epsfig{file=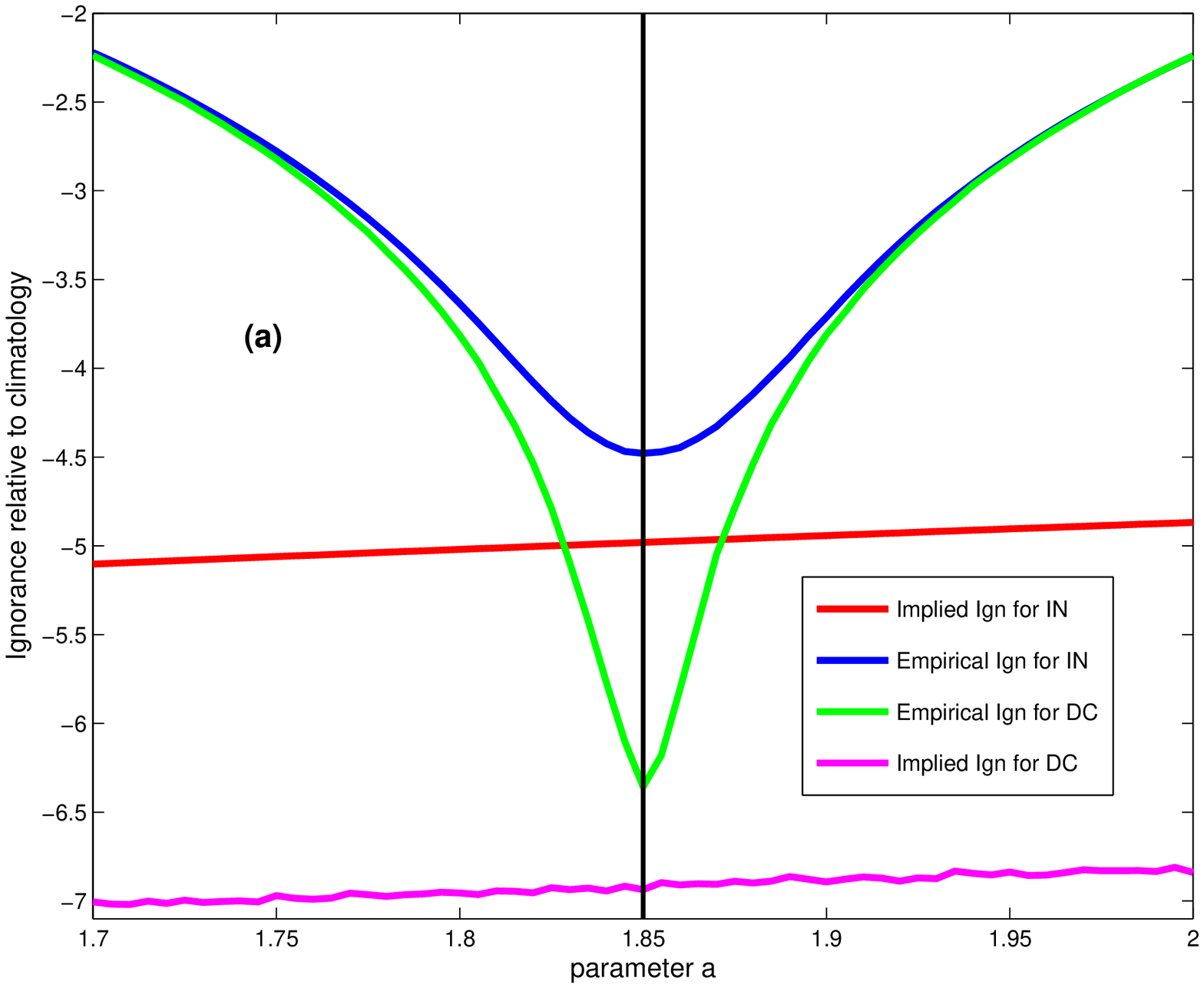, width=6.5cm}
  \epsfig{file=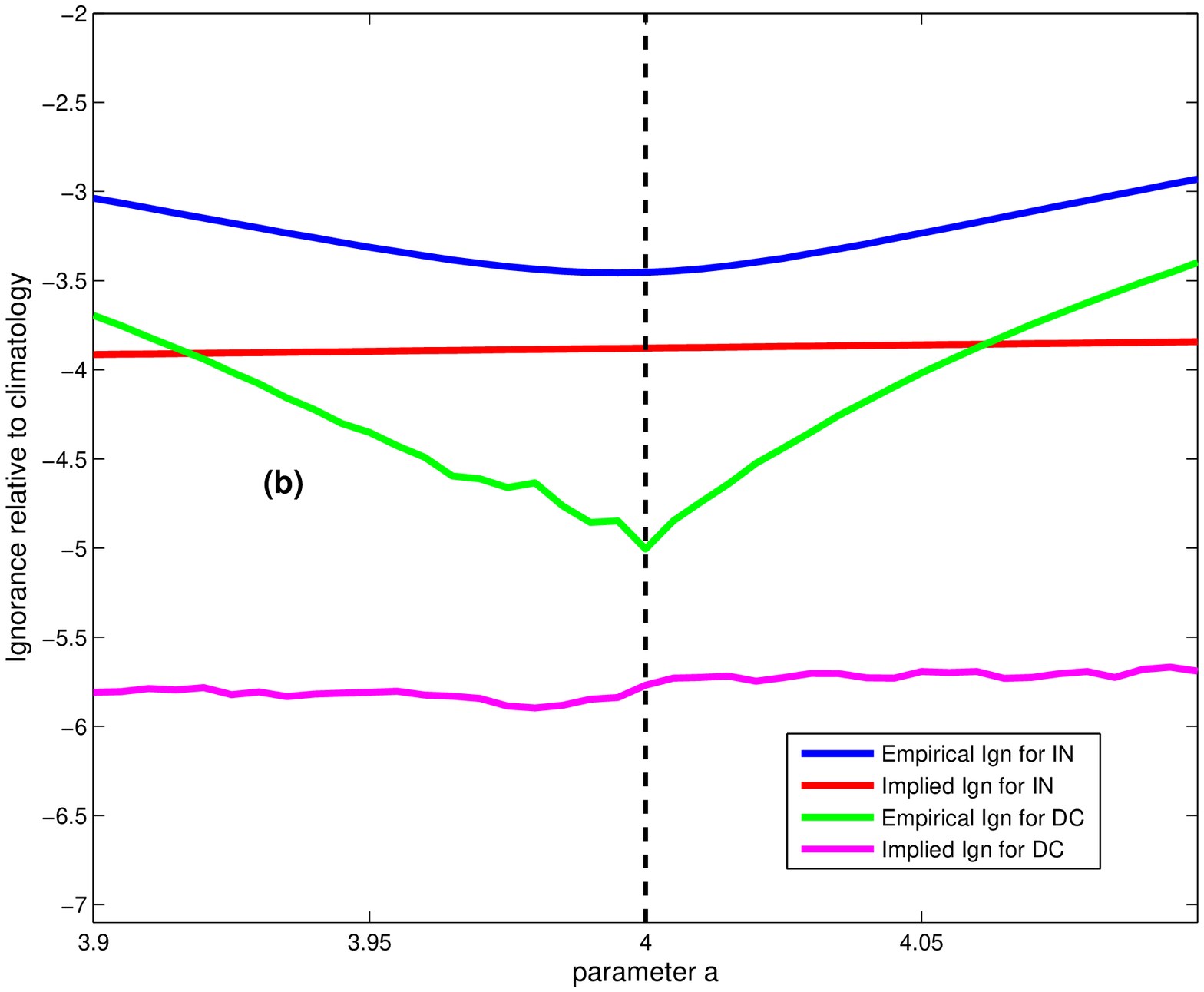, width=6.5cm}
}
\vbox{
  \epsfig{file=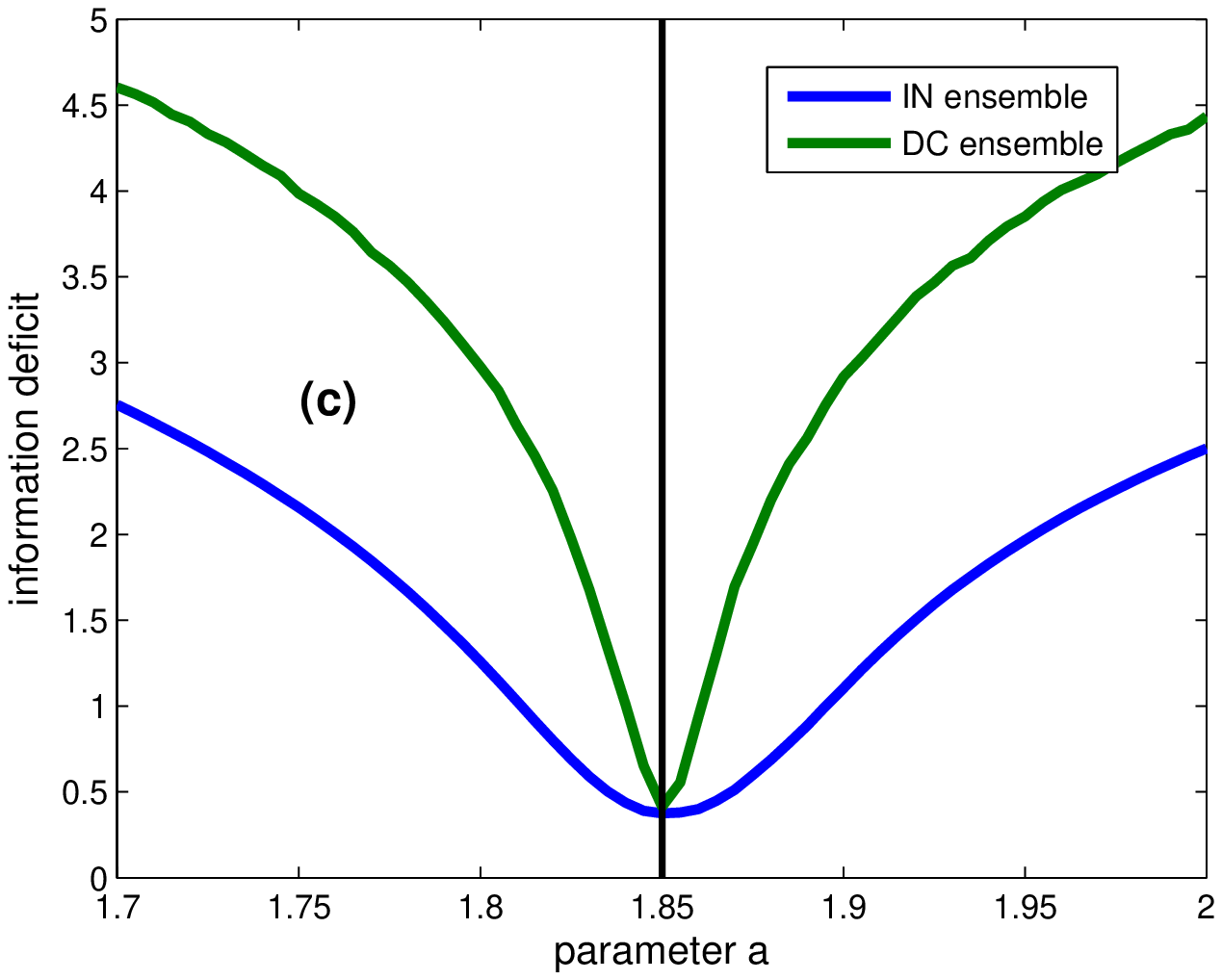, width=6.5cm}
  \epsfig{file=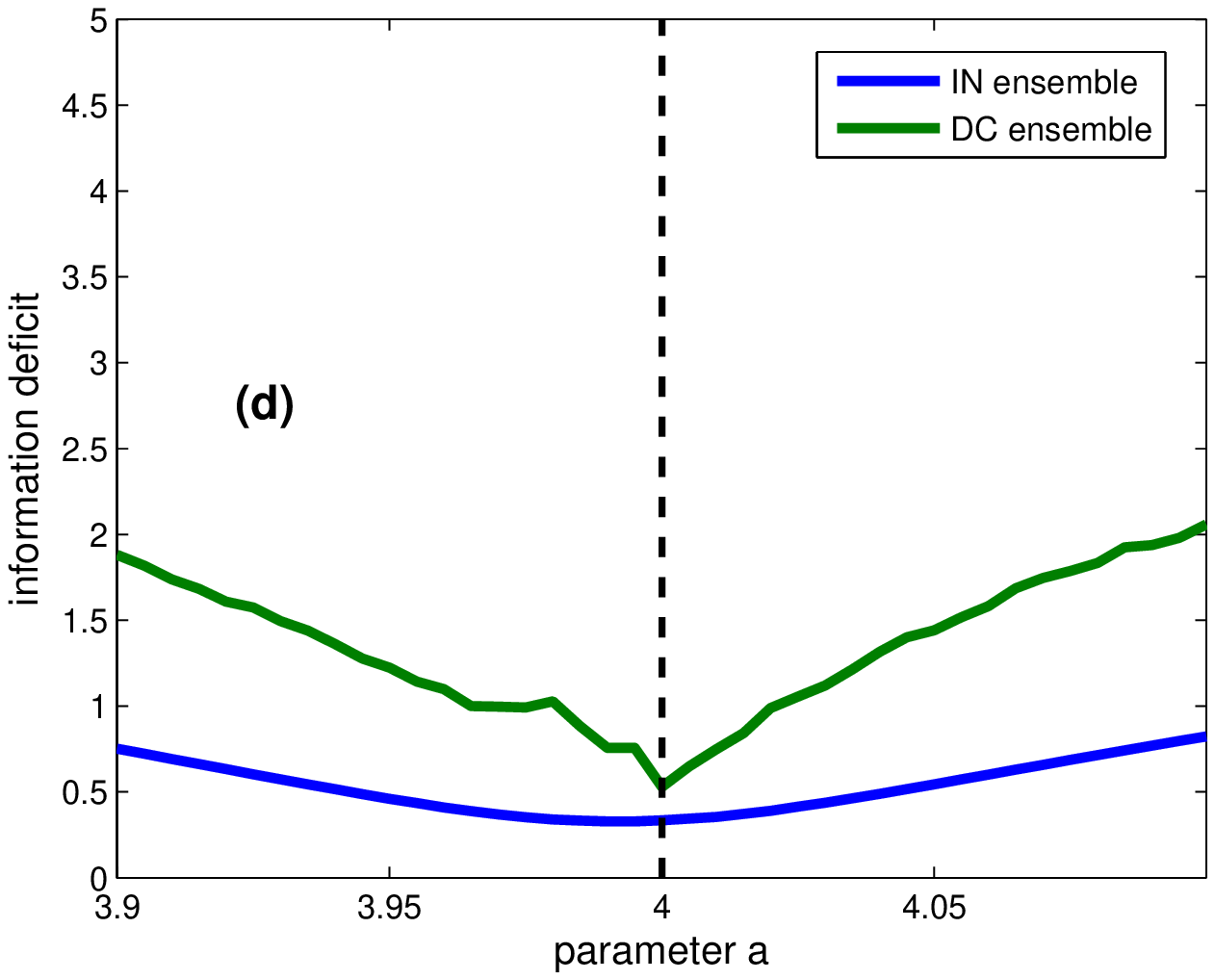, width=6.5cm}
}
  \caption{(Color) Empirical Ignorance and Implied Ignorance as a function of parameter value with noise level $\sigma=1/128$ for lead time 1. Curves for both Inverse Noise ensemble and Dynamically consistent ensemble. 1024 forecasts are considered in each case. (a) the Perfect Model Scenario with the Logistic Map: $F(x,a)=1-ax^2$, (b) the Imperfect Model Scenario with system-model pair of Equation~\ref{eq:qsystem}~\&~\ref{eq:logisticMap}, (c) information deficit in the Perfect Model Scenario, (d) information deficit in the Imperfect Model Scenario.}
  \label{fig:infodefic}
\end{figure}

``Potential predictability'' reflects the utility an existing forecast system would have {\it if} it were perfect~\cite{glossary}. Interpreting this as utility carries some risk, of course as the actual system may be much more predictable (or much less) than the dynamics of the current generation of models. An alternative approach which can quantify the (historical) impact of model inadequacy is to contrast the Empirical Ignorance with the Implied Ignorance, defined as
\begin{eqnarray}
 \label{eq:expign}
    \int -p_{m}(y)log_{2}(p_{m}(y)) dy
\end{eqnarray}
The Implied Ignorance is the Ignorance one would expect to observe if in fact the probability forecast was perfect. The difference between Empirical Ignorance and Implied Ignorance reveals an {\it information deficit} (in bits), which exposes shortcomings anywhere in the forecast methodology. In contrast with the so called ``estimate'' of skill from ``potential predictability'' experiments which assumes the model is perfect, the information deficit quantifies just how far the predictability of the current model is from (its internal) perfection. Within the Perfect Model Scenario, the Implied Ignorance can approach the Empirical Ignorance for the ``True'' parameter values (if and only if the entire ensemble forecasting package is perfect). Even when the model structure is mathematically correct, the Empirical Ignorance may be greater than the Implied Ignorance, indicating that the model PDF is an incomplete reflection of the expected uncertainties.\footref{fn:foot} Indeed the information deficit provides quantitative information on second order uncertainty. Figure~\ref{fig:infodefic} illustrates this in both the perfect model case (Figure~\ref{fig:infodefic}(a) and (c)) and the imperfect model case (Figure~\ref{fig:infodefic}(b) and (d)). In the perfect model case, the Empirical Ignorance and Implied Ignorance should converge to within sampling error at the ``True'' parameter when a (many step) dynamically consistent ensemble is employed. In Figure~\ref{fig:infodefic}(a) and (b), the upper blue line shows the Empirical Ignorance for Inverse Noise ensemble, the lower green line shows the Empirical Ignorance for a (one step) dynamically consistent ensemble. The upper red line and lower purple line correspond to the Implied Ignorance for each ensemble formation strategy respectively. Figure~\ref{fig:infodefic}(c) and (d) show the information deficits correspond to Figure~\ref{fig:infodefic}(a) and (b). The information deficit will remain nonzero as long as Inverse Noise is used. In Figure~\ref{fig:infodefic}(a) and (c) the information deficit for dynamically consistent ensembles remains nonzero because these dynamically consistent ensembles are only consistent with one observation, as the window of dynamically consistency increases and the ensemble size increases the information deficit will approach zero. On the other hand, in the imperfect model case (Figure~\ref{fig:infodefic}(b) and (d)) the information deficit will remain nonzero no matter what one may do due to the model inadequacy.

Also note in Figure~\ref{fig:infodefic} that the information deficit of Inverse Noise ensemble is smaller than that of dynamically consistent ensemble. This is somewhat misleading, in the same way that potential predictability is consistently misleading. Confusion can be avoided by noting that forecasts using the (one step) dynamically consistent ensemble provide almost 2 bits more information beyond those from the Inverse Noise ensembles.

\begin{center}
{\bf Conclusion}
\end{center}
Although widely popular, the method of LS is optimal only in a narrow context, a fact stressed by Kalman~\cite{Kalman}; LS is often applied well outside its mathematical remit. While a general account of parameter estimation remains lacking, the straightforward Minimum Ignorance approach introduced here is shown to yield good parameter estimation in several chaotic systems.
Initial experiments suggest that the MI approach is also useful
for identifying best parameter in an imperfect model as long as the notion of ``best'' is well defined.

MI is expected to perform well against the myriad of modern alternatives. The initial value approach~\cite{Schittkowski} is reminiscent of four-dimensional variational assimilation~\cite{Lorenc} (4DVAR), minimizing the cost function not only for the initial condition but also for the parameter values; like 4DVAR it is computationally expensive and suffers from local minima. These variational methods differ from the LS method, inasmuch as LS fails fundamentally while the initial value approach fails numerically. Simply put, the root of this failure lies in chaotic likelihoods~\cite{Berliner}. Voss~\cite{Voss} applied a multiple shooting method to address the local minima problem in initial value approach; an initial value approach in short windows, resembling a similar spin up procedure applied to 4DVAR~\cite{Talagrand}; the approach remains expensive and Voss's examples show varying success. MI might be considered as a useful pre-filter for the method of~\cite{Voss}; even then that method requires ad hoc continuity constraints.

Sequential (recursive) methods provide an alternative approach. Kalman filter methods are most often applied to state estimation, to estimate the parameter one may simply add the parameter vector to the state vector~\cite{Voss}.
For weakly nonlinear systems the extended Kalman Filter~\cite{Ghil} can be used. For strong nonlinear systems, Voss~\cite{Voss} introduced the unscented Kalman filter. Notice that for such of sequential methods, the parameter vector is allowed to evolve in time; Expectation Maximization (EM) algorithms~\cite{Dempster} account for this. Each of these methods perform better where the Gaussianity assumption holds more strongly. MI does \textbf{not} require any Gaussian constraints what so ever.

MI parameter estimation also has the advantage that it is easy to use. Methods which contrast the natural measure of the model with the observations~\cite{mcsharrys99} are significantly more complicated, and grow more so as the dimensionality of the model increases. Alternative methods which contrast shadowing times of the model as a function of parameter values~\cite{IS3} are significantly more computationally expensive. MI estimation using Inverse Noise ensembles is straightforward to implement and relatively inexpensive computationally. It will fail to indicate the ``True'' parameter value when the ensemble is not distributed consistently with respect to the model's long term dynamics (natural measure), but the parameter value MI suggests will give better probabilistic  forecasts than the ``True'' parameter value {\bf as long as} the flawed ensemble formation scheme is used. Investing more in data assimilation is shown to improve parameter estimates. When the mathematical structure of the model is incommensurate with the structure of the system generating the observations, the ultimate goal of parameter estimation is unclear. As illustrated above, the ``optimal'' parameter value may, for example, vary with lead time. In such cases, MI can still provide useful parameter estimation as long as the goal (``optimal'') is well defined.

MI parameter estimation by ensemble prediction provides a useful new tool, avoiding the shortcomings of other approaches. The information deficit reflected in the Implied Ignorance can reveal forecast system inadequacies and quantify the predictability in a more informative manner than ``potential predictability'' does. We are optimistic that this framework will allow some progress outside the Perfect Model Scenario.

\begin{center}
{\bf Acknowledgment}
\end{center}

This research was supported both by the LSE's Grantham Research Institute and the ESRC Centre for Climate Change Economics and Policy, funded by the Economic and Social Research Council and Munich Re. L.A.S. gratefully acknowledges support from Pembroke College, Oxford.


\begin{thebibliography}{990}

\bibitem{Sornette}
V.F. Pisarenko and D. Sornette, Statistical methods of parameter estimation for deterministically chaotic time series. {\it Phys. Rev. E}, {\bf 69}, (2004).

\bibitem{Stigler}
S.M. Stigler, Gauss and the invention of least squares. {\it Ann. Stat.} {\bf 9}(3), 465, (1981).

\bibitem{Teets}
D. Teets and K. Whitehead, The Discovery of Ceres: How Gauss Became Famous. {\it Mathematics Magazine}, {\bf 72}, 83-91, (1999).

\bibitem{Gauss}
C.F. Gauss, Theoria Motus. C H Davis (translator) Dover Publications, New York, (1857).

\bibitem{Stern}
L.A. Smith and N. Stern, Uncertainty in Science and its Role in Climate Policy. in press for {\it Royal Society Phil Trans A}, (2011).

\bibitem{Chaosbook}
L.A. Smith, A Very Short Introduction to Chaos. Oxford University Press, (2007).

\bibitem{Smith02}
L.A. Smith, What Might We Learn from Climate Forecasts? {\it Proc. National Acad. Sci.} {\bf 4}(99), 2487-2492, (2002).

\bibitem{mcsharrys99}
P.E. McSharry and L.A. Smith, Better nonlinear models from noisy data: Attractors with maximum likelihood. {\it Phys. Rev. Lett.} {\bf 83}, 4285, (1999).

\bibitem{Crevel}
D.R. Creveling, P.E. Gill and H.D.I. Abarbanel, State and Parameter Estimation in Nonlinear Systems as an Optimal Tracking Problem. {\it Phys. Lett. A}, {\bf 372}, 2640, (2008).

\bibitem{Maybhate}
A. Maybhate and R.E. Amritkar, Use of synchronization and adaptive control in parameter
estimation from a time series. {\it PHYSICAL REVIEWE},{\bf 59}, 284-293, (1999).

\bibitem{IS3}
L.A. Smith, M.C. Cuellar, H. Du and K. Judd, Exploiting dynamical coherence: A geometric approach to parameter estimation in nonlinear models. {\it Phys. Lett. A}, {\bf 374}, 2618, (2010).

\bibitem{Heald}
J. Heald and J. Stark, Estimation of noise levels for models of chaotic dynamical systems. {\it Phys. Rev. Lett.}, {\bf 84} (11), 2366, (2000).

\bibitem{Abarbanel}
J.C. Quinn and Henry D.I Abarbanel, State and parameter estimation using Monte Carlo evaluation of path integrals. {\it Quarterly Journal of the Royal Meteorological Society}, {\bf 136}, 1855, (2010).

\bibitem{Schittkowski}
K. Schittkowski, Parameter estimation in systems of nonlinear equations, {\it Numerische Mathematik}, {\bf 68}, 129-142 (1994).

\bibitem{Voss}
H.U. Voss, J. Timmer,  and J. Kurths, Nonlinear Dynamical System Identification from Uncertain and Indirect Measurements, {\it I. J. Bifurcation and Chaos}, 1905-1933, (2004).

\bibitem{Dempster}
A. P. Dempster, N. M. Laird and D. B. Rubin, Maximum Likelihood from Incomplete Data via the EM Algorithm, {\it Journal of the Royal Statistical Society}, {\bf 39}(1), 1¨C38, (1977).

\bibitem{Ghahramani}
Z. Ghahramani and S. Roweis, Learning nonlinear dynamical systems using an EM algorithm, Advances in Neural Information Processing Systems 11, MIT Press, 431-437, (1999).

\bibitem{Ghil}
M. Ghil and P. Malanotte-Rizzoli, Data assimilation in meteorology and oceanography, {\it Adv. Geophys.}, {\bf 33}, 141-266, (1991).

\bibitem{Bos}
A. Van Den Bos, Parameter estimation for scientists and engineers. Wiley-Interscience, (2007).

\bibitem{Bollt}
C. Yao, E.M. Bollt, Modeling and Nonlinear Parameter Estimation with Kronecker product Representation for Coupled Oscillators and Spatiotemporal Systems. {\it Physica D} {\bf 227} 1, 78, (2007).

\bibitem{LSE}
A. Bjorck, Numerical Methods for Least Squares Problems. Philadelphia, SIAM, (1996).

\bibitem{Inference}
G. Casella and R.L. Berger, Statistical Inference, Wadsworth and Brooks Cole, California, (1990).

\bibitem{TLSE}
S. Van huffel and J. Vandewalle, The Total Least Squares Problem. Philadelphia, SIAM, (1991).

\bibitem{Taran}
A. Tarantola, Inverse Problem Theory and Model Parameter Estimation. Philadelphia, SIAM, (2004).

\bibitem{Lorenz96}
E.N. Lorenz, Predictability: A problem partly solved. ECMWF Seminar Proceedings on Predictability. Reading, UK, (1995).

\bibitem{Lorenz65}
E.N. Lorenz, A study of the predictability of a 28-variable atmospheric model. {\it Tellus}, {\bf 17}, 321, (1965).

\bibitem{Fermi}
L.A. Smith, Proc International School of Physics ``Enrico Fermi'', CXXXIII, 177, Bologna, Italy (1997).

\bibitem{Adress}
J. Brocker and L.A. Smith, From Ensemble Forecasts to Predictive Distribution Function. {\it Tellus A}, {\bf 60}, 663, (2008).

\bibitem{Good}
I.J. Good, Rational decisions. {\it J. R. Stat. Soc.}, Series B {\bf XIV} (1) 107, (1952).

\bibitem{IgnRS}
M.S. Roulston and L.A. Smith, Evaluating probabilistic forecasts using information theory. {\it Mon. Weather Rev.}, {\bf 130}, 1653, (2002).

\bibitem{proper}
J.M. Bernardo, Expected information as expected utility. {\it Ann. Stat.}, {\bf 7}, 686, (1979).

\bibitem{Score}
J. Brocker, L.A. Smith, Scoring Probabilistic Forecasts: On the Importance of Being Proper. {\it Weather Forecasting}, {\bf 22} (2), 382, (2006).

\bibitem{EnKF94}
G. Evensen, Sequential data assimilation with a nonlinear quasigeostrophic model using Monte-Carlo methods to forecast error statistics. {\it J. Geophys. Res.}, {\bf 99(C5)}, 10 143, (1994).

\bibitem{IS1}
K. Judd and L.A. Smith, Indistinguishable States I: The Perfect Model Scenario. {\it Physica D}, {\bf 151}, 125, (2001).

\bibitem{Palmer}
M. Leutbecher and T.N. Palmer, Ensemble forecasting. {\it J. Comp. Phys.} {\bf 227} 3515, (2008).

\bibitem{Raftery}
A.E. Raftery, et al, Using Bayesian Model Averaging to Calibrate Forecast Ensembles. {\it Mon. Weather Rev.}, {\bf 131}, 1155, (2005).

\bibitem{May}
R.M. May, Simple mathematical models with very complicated dynamics. {\it Nature} 261, (1976).

\bibitem{Wilks06}
D.S. Wilks, Comparison of ensemble-MOS methods in the Lorenz `96 setting. {\it Meteorol. Appl.}, {\bf 13}, 243, (2006).

\bibitem{Farmer}
J.D. Farmer and J.J. Sidorowich, Predicting chaotic time series. {\it Phys. Rev. Lett.}, {\bf 59}, 845, (1987).

\bibitem{Bollt07}
E. M. Bollt, Attractor Modeling and Empirical Nonlinear Model Reduction of dissipative Dynamical Systems. {\it I. J. Bifurcation and Chaos} {\bf 17(4)}, 1199-1219, (2007).

\bibitem{glossary}
AMS (American Meteorological Society), Glossary of Meteorology. T. S. Glickman, ed., 2nd Edition. Boston: American Meteorological Society, (2000).

\bibitem{Kalman}
R.E. Kalman, A New Approach to Linear Filtering and Prediction Problems. {\it Trans AMSE Ser. D J Basi Eng}, {\bf 82}, 35, (1960).

\bibitem{Lorenc}
A.C. Lorenc, Analysis methods for numerical weather prediction, {\it Quart. J. Roy. Meteor. Soc.}, {\bf 112},1177-1194,(1986).

\bibitem{Berliner}
M.L. Berliner, Likelihood and Bayesian Prediction for Chaotic Systems, {\it J. Am. Stat. Assoc.}, {\bf 86}, 938-952, (1991).

\bibitem{Talagrand}
C. Pires, R. Vautard, and O. Talagrand, On extending the limits of variational assimilation in nonlinear chaotic systems. {\it Tellus}, {\bf 48A}, 96-
121, (1996).

\end{thebibliography}
\end{document}